\documentclass[twocolumn,showpacs,prl]{revtex4}
\usepackage{amsmath,amsthm,amscd}
\usepackage{dcolumn}
\usepackage{graphicx}
\usepackage{epsfig}
\begin{document}
\newcommand{\Qb}{\ensuremath{\mathbf{Q}}}
\newcommand{\qb}{\ensuremath{\mathbf{q}}}
\newcommand{\Sb}{\ensuremath{\mathbf{S}}}
\newcommand{\nb}{\ensuremath{\mathbf{n}}}
\newcommand{\mb}{\ensuremath{\mathbf{m}}}
\newcommand{\ub}{\ensuremath{\mathbf{u}}}
\newcommand{\vb}{\ensuremath{\mathbf{v}}}
\newcommand{\xb}{\ensuremath{\mathbf{x}}}
\newcommand{\yb}{\ensuremath{\mathbf{y}}}
\title{Frustrated classical Heisenberg and XY models in 2 dimensions with
nearest-neighbor biquadratic exchange: exact solution for the ground-state phase diagram}
\author{L. X. Hayden$^1$, T. A. Kaplan$^2$, and S. D. Mahanti$^2$}
\affiliation{$^1$Department of Physics, University of Missouri, Columbia, MO\\
$^2$Department of Physics \& Astronomy and Institute for Quantum Sciences, Michigan State University, East
Lansing, MI 48824}
\begin{abstract}
The ground state phase diagram is determined exactly for the frustrated classical Heisenberg model plus
nearest-neighbor biquadratic exchange interactions on a 2-dimensional lattice. A square- and a rhombic-symmetry
version are considered. There appear ferromagnetic, incommensurate-spiral, ``up-up-down-down" (uudd) and canted
ferromagnetic states, a non-spiral coplanar state that is an ordered vortex lattice, plus a non-coplanar ordered
state (a ``conical vortex lattice"). In the rhombic case, which adds biquadratic terms to the Heisenberg model
used widely for insulating manganites, the uudd state found is the E-type state observed; this along with
accounting essentially for the variety of ground states observed in these materials, shows that this model
probably contains the long-sought mechanism behind the uudd state.
\end{abstract} 
\maketitle

\textbf{I. Introduction}\\
A classical spin model studied by Thorpe and Blume~\cite{thorpe} (TB) showed interesting ground state behavior,
where there was either simple collinear-spin long range order, or disorder. The spins were on a linear chain,
with nearest-neighbor (nn) Heisenberg and biquadratic exchange interactions. Recently a next-nearest-neighbor
(nnn) anti-ferromagnetic Heisenberg exchange term was added (making the Heisenberg terms frustrated), solved
exactly for the ground state, and found to yield a rich phase diagram,~\cite{kaplan} with spirals and the
``up-up-down-down" (uudd) state (isotropic version of the uudd state of the ANNNI model~\cite{fisher}), plus the
TB states.

It was speculated~\cite{kaplan} that extension of the model to lattice dimensionality d = 2, with the rhombic
symmetry of the Heisenberg model used for multiferroic manganites~\cite{kimura,kaplan2,mochizuki}, would yield
the historically puzzling uudd (E-type) state observed in those materials.

Here we carry out this extension, and also treat a corresponding square-symmetry model. We again find the ground
state exactly. As in~\cite{kaplan}, this is enabled by use of the LK cluster method~\cite{lyons}; it is also an
additional test of the applicability of that method.

A 2d version of the uudd state is indeed found in the rhombic model and is essentially the observed uudd
state~\cite{kimura,zhou}. Spirals and highly degenerate phases are also found. A model along these lines appears
to be realistic for the manganites, and provides strong support for the suggested mechanism [2] behind the uudd
state, namely frustrated Heisenberg plus biquadratic interactions.

For the square symmetry, a coplanar non-spiral state that is an ordered array of vortices, a ``vortex lattice"
(VL), is found, also discussed earlier by Henley~\cite{henley2} (see also~\cite{chubukov}), both for XY and
Heisenberg spins. Also found is a non-coplanar state, a ``conical vortex lattice".

A principal motivation for the addition of biquadratic terms to the frustrated Heisenberg model~\cite{kaplan}
was that they can be large for ions with large spin $S$.~\cite{harris, rodbell} Two sources of these terms are
i. Electronic: higher order terms in the hopping amplitudes or orbital overlap (leading order yields the
Heisenberg interactions)\cite{anderson, huang} and ii. Lattice induced via spin-lattice
interaction~\cite{kittel,wagner}. There are indications that these sources may be of roughly equal
magnitude.~\cite{harris, rodbell,anderson,huang}. For the present purposes, the source is not relevant.

%In fact for Mn$^{3+}$, $S=2$, relevant to the manganites, these terms are expected to be $>>$ than the usual
%anisotropic terms in previous models of magnetic ordering in the insulating manganites (e.g.~\cite{mochizuki}),
% 5o they must be considered (see further discussion below). However, adding only biquadratic terms is
%phenomenological: Indications are that the main mechanism for such terms is higher-order in the hopping a
%amplitudes,~\cite{rodbell} which in leading order yield the Heisenberg terms.~\cite{anderson} But there are
%additional terms in the same order (($t/U)^2$ times the Heisenberg terms, in Hubbard model language), namely
%3-~\cite{bastardis} and 4-body~\cite{takahashi} terms. Thus this work is in the spirit of exploring simple
%isotropic terms beyond Heisenberg.

The model Hamiltonian studied is
\begin{eqnarray}
H&=&\sum_{<\nb,\mb>}[J_1\Sb_\nb\cdot\Sb_{\mb}-A(\Sb_\nb\cdot\Sb_{\mb})^2]\nonumber\\
&    &\mbox{}+ J_2\sum^1_{<\nb,\mb>}\Sb_\nb\cdot\Sb_{\mb}+J_2^\prime\sum^2_{<\nb,\mb>}\Sb_\nb\cdot\Sb_{\mb},
\end{eqnarray}
where $\Sb_\ub$, a unit 3-vector, is the spin at site $\ub$. The first term sums Heisenberg and biquadratic
interactions over nn pairs: $\nb,\mb$ go over the vectors of a square lattice. The 2nd and 3rd terms are,
respectively, sums over the nn pairs along the (1,1) and (1,-1) diagonals of the square unit cell. We consider
two cases: $J_2=J_2^\prime$ (square symmetry) and infinitesimal $J_2^\prime$ (rhombic symmetry). The latter case
is motivated by models~\cite{kimura,kaplan2,mochizuki} applied to manganites.\cite{ref}

$H$ extends that studied in~\cite{kaplan} to d = 2. Motivations for its study are as in ~\cite{kaplan}, e.g.
biquadratic terms can be large for large-spin ions~\cite{harris,rodbell}, such terms are used to mimic the
order-selecting effects of thermal, quantum, or dilution fluctuations ( ``order-by-disorder"
effects)~\cite{henley,nikuni}, its ground state phase diagram can be found analytically, and shows properties
that should be of interest in statistical mechanics and for manganites particularly.

The Luttinger-Tisza method and its generalizations (see the review~\cite{kaplan3}) appear to be not useful in
connection with (1) because of the non-linearity in the equation for stationarity of $H$ subject to the weak
constraint, $\sum_j(J_{ij}-2A_{ij}\Sb_i\cdot\Sb_j)\Sb_j=\lambda\Sb_i.$

Instead we turn to the rather unknown LK cluster method~\cite{lyons}, which solves the problem exactly. Recall
that method as applied here. Assume periodic boundary conditions, with the thermodynamic limit (TL) to be taken
finally.\cite{lyons} Then (1) can be written
\begin{equation}
H=\sum_\nb H_c(\Sb_\nb,\Sb_{\nb+\hat{x}},\Sb_{\nb+\hat{x}+\hat{y}},\Sb_{\nb+\hat{y}}),\label{2}
\end{equation}
where $H_c$ is the cluster energy; $h_c\equiv H_c/|J_1|$ is given by
\begin{eqnarray}
h_c(\Sb_1,\Sb_2,\Sb_3,\Sb_4)&=&-\frac{1}{2}\sum_{n=1}^4[\Sb_n\cdot\Sb_{n+1}+a(\Sb_n\cdot\Sb_{n+1})^2]\nonumber\\
&  &+\gamma\Sb_1\cdot\Sb_3+\gamma^\prime\Sb_2\cdot\Sb_4,\label{3}
\end{eqnarray}
where $\Sb_5\equiv\Sb_1$, $a=A/|J_1|, \gamma=J_2/|J_1|,\gamma^\prime=J_2^\prime/|J_1|$, and we've taken $J_1<0$.
Clearly, $h\equiv H/|J_1|$ satisfies
\begin{equation}
h\ge \sum_\nb \min h_c(\Sb_\nb,\Sb_{\nb+\hat{x}},\Sb_{\nb+\hat{x}+\hat{y}},\Sb_{\nb+\hat{y}}).\label{4}
\end{equation}
If states that minimize $h_c$ "propagate", i.e. if there is a state of the whole system such that every cluster
(every square plaquette with its 4 spins) achieves the minimum $h_c$, it follows that the state is a ground
state of $H$ (the global minimum). To minimize $h_c$, we find, analytically, stationary states, construct a
phase diagram by comparing their $h_c-$values and check that there are no lower states by calculating $h_c$ on a
mesh over the whole range of the variables. This and other related matters are discussed in the Appendix.
\begin{figure}[h]
\includegraphics[height=1.7in]{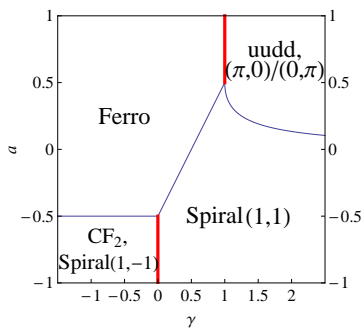}
 \caption{(Color online) Phase diagram,  $\gamma^\prime=0$ (rhombic symmetry). In the upper-right and lower-left regions
 there is large degeneracy that is lifted by $\gamma^\prime\ne0$ in favor of the states given.}
 \label{fig:phase diagram1}
\end{figure}

\textbf{II. Results.} See Appendix for their derivations.\\
\textbf{Case 1. Infinitesimal $\gamma^\prime$ (rhombic symmetry)}\\
For clarity, we first consider coplanar spins (spin dimensionality D=2, i.e. XY spins). Because of the
spin-isotropy of $h_c$, it is only a function of 3 angles. FIG. 1 is the phase diagram. The state, all spins
parallel, occurs in the Ferro region. In the upper-right region, the states $\mbox{uudd or } (\pi,0)/(0,\pi)$,
shown in FIG.2, are the ground states for $\gamma^\prime < \mbox{or} > 0$;
\begin{figure}[h]
\includegraphics[height=1.3in]{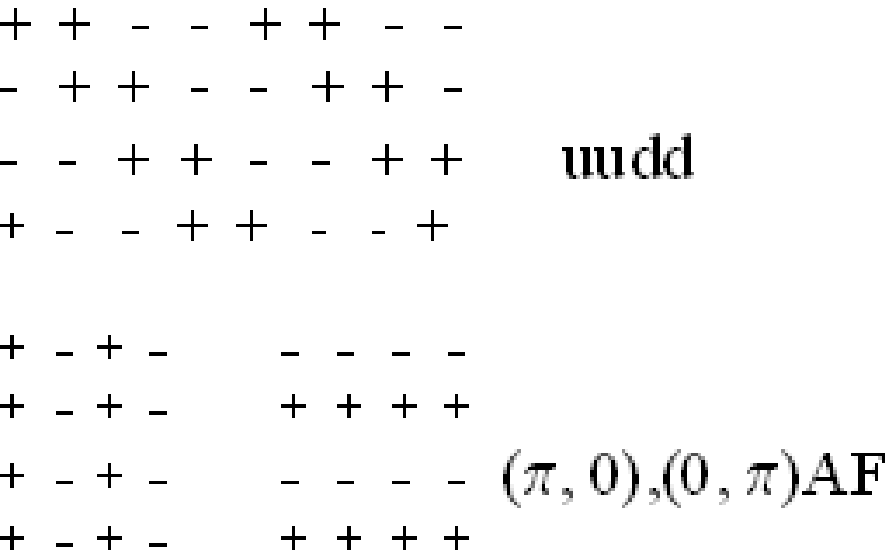}
 \caption{The ground states in the $\mbox{uudd},(\pi,0)/(0,\pi)$ region of FIG. 1.}
 \label{fig:some ground states}
\end{figure}
$(\pi,0)$ and $(0,\pi)$ refer to propagation vectors. The uudd state is a wave with propagation vector $\qb$ in
the (1,1) direction. The notation is $(q_x,q_y)$, x-axis to the right, y up.

In the Spiral region is a simple spiral~\cite{kaplan3} with propagation vector $\qb=(q_0,q_0)$,
 $\cos q_0=[2(\gamma-a)]^{-1}$.

\begin{figure}[h]
\includegraphics[height=1.5in]{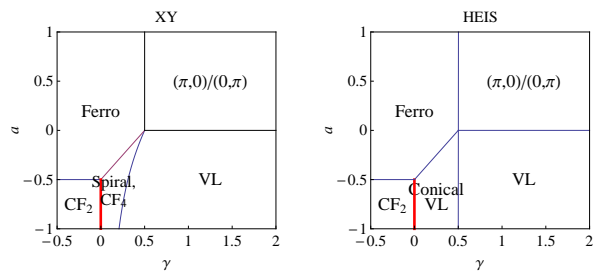}
 \caption{(Color online) Phase diagrams, $\gamma^\prime=\gamma$ (square symmetry), for XY and Heisenberg (HEIS)
 models, respectively}
 \label{phase diagram2}
\end{figure}

In the lower left region a canted ferromagnet, CF$_2$, shown in Fig. 4, or a spiral are ground states for
$\gamma^\prime < \mbox{or} > 0$. The spiral wave vector is $(q_1,-q_1), \cos q_1=-1/(2a)$, $q_1$ being also the
canting angle.

The phase diagram is unchanged for Heisenberg spins.\\
\textbf{Case 2. $\gamma=\gamma^\prime$ (square symmetry)}\\
FIG. 3 shows the phase diagrams for XY and for Heisenberg spins. \textbf{XY}: The Ferro region is similar to
that in FIG.1. The $(\pi,0),(0,\pi)$ states no longer coexist with the uudd states ($\gamma>1/2, a>0$). The
ground state in the VL region, discussed previously by Henley~\cite{henley2} (who considered only $\gamma>1/2$),
can be described as an ordered array of vortices, which we call a vortex lattice. See FIG. 4 for an example,
where the filled and unfilled circles indicate a pair of vortices of opposite vorticity. The vortices form a
square lattice. In the region labelled Spiral, CF$_4$, a $(q_0,q_0)$ spiral and a canted ferromagnet, CF$_4$
(see Fig. 4) are degenerate ground states. In the extreme lower left, the ground state CF$_2$ is no longer
degenerate with a spiral. This canted ferromagnet was also found in~\cite{chubukov}. \textbf{HEIS}: The main
change from XY to HEIS is the replacement of the Spiral-CF$_4$ phase by  a non-
coplanar state, discussed below. \\
\begin{figure}[h]
\includegraphics[height=1.2in]{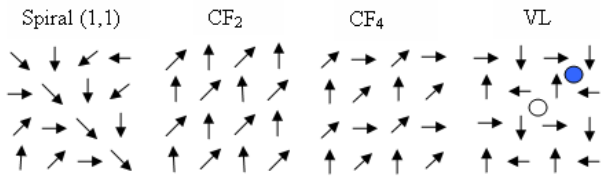}
 \caption{(Color online) Spiral and canted ferromagnets, CF$_n$ (for illustrative value $\pi/4$ of the
 turn-angle $q_0$.). Vortex lattice: ground state in regions VL of FIG. 3.}
 \label{fig:XY-spin ground states}
\end{figure}

\textbf{Non-coplanar states}\\
    We found the ground state to be non-coplanar in the region Conical VL (FIG. 3HEIS). FIG. 5 shows an example.
There appears no obvious symmetry, although it was found that at all points in the region, $\theta_2=\theta_4$
and $\phi_3=(1/2)\phi_4$. After FIG. 5 was drawn, and much puzzlement, we found that a particular uniform
rotation of the spins brings the state to a highly symmetric one: The spins in each plaquette lie on the surface
of a cone, of half-angle $\Omega$,  and the azimuthal angles are equally spaced (i.e. the spacing is $\pi/2$).
Thus the name "Conical VL". $\Omega$ varies smoothly from 0 at the Ferro boundary to $\pi/2$ at the VL boundary.
But at the CF$_2$ boundary
there is a first-order transition.  Note that there is a net spin, i.e. this is ferro- (or ferri-) magnetic.\\

\textbf{Degeneracies}\\
In classical systems variables vary continuously. However, in the XY case, fixing just one spin in our ground
states makes them countable: They derive from various propagations of the degenerate cluster ground states,
which are clearly countable when one spin is fixed. This allows the definition of entropy $\mathcal{S}= \ln$
(number of states), which we will use for XY spins.

In the CF$_2$ and uudd regions of FIG.1 there is a large degeneracy coming from many ways of propagating the
cluster ground states: the corresponding entropy $\mathcal{S}\ge N^{1/2}\ln2$, where $N$ is the number of spins.
Non-zero $\gamma^\prime$ removes this degeneracy. In the Spiral-CF$_4$ region of FIG.3XY there is a similarly
large degeneracy.

The propagation of the Ferro and Spiral states, FIG. 1, is unique; but we cannot conclude they are
non-degenerate~(see Appendix). Similarly, all the regions in FIG. 3 other than Spiral-CF$_4$ show unique
propagation.

The emphasized line segments at $\gamma=0$ and 1 in FIG. 1 and at $\gamma=0$ in FIG. 3 are closely related to
the disorder lines in the 1d case~\cite{kaplan}. The 2d generalization of the TB disordered states~\cite{thorpe}
occurs at $\gamma=0$. In 1d, $\mathcal{S}=N\ln2$. Whether a similar conclusion holds in 2d is an interesting
question that should be addressed. We find $\mathcal{S}$ is at least O($N^{1/2})$~(see Appendix) The line at
$\gamma=1$, FIG.1, is the 2d isotropic generalization of the highly degenerate states of the ANNNI
model~\cite{fisher} at the multiphase point.

\begin{figure}[h]
\includegraphics[height=1.1in]{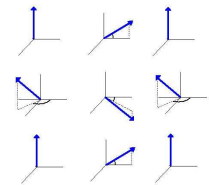}
 \caption{(Color online) Non-coplanar ground state in the Non-coplanar region of FIG. 3HEIS at
  (a,$\gamma)$=(-0.5,0.3).
 $\theta_2=\theta_4=66.42^o,\theta_3=101.5^o,\phi_3=(1/2)\phi_4=57.7^o$. }
 \label{non-coplanar state}
\end{figure}

\textbf{III. Discussion}

\textbf{Case 1. $\gamma^\prime=0$, extreme rhombic symmetry}. The speculation~\cite{kaplan} that the d=2 version
of the rhombic model would be qualitatively similar to the d=1 case, is borne out: the phase diagram FIG. 1 is
topologically the same as that for d=1~\cite{kaplan}. There are however three major differences. The Ferro-uudd
boundary occurs at $\gamma=1$ for d = 2, vs. $\gamma=1/2$ for d=1. While the uudd state is the only state in its
region for d=1, in d=2 there are the other degenerate states, $(\pi,0),(0,\pi)$. Similarly, in 1d the CF$_2$
state appears alone in its region, while in 2d it is degenerate with a (1,-1) spiral.

Experimentally it is uudd, not $(\pi,0),(0,\pi)$, that is observed~\cite{kimura,zhou}. As seen from FIG. 2, a
small $\gamma^\prime$ will remove that degeneracy, a \emph{ferromagnetic} $\gamma^\prime$ will favor the uudd
state. Interestingly, the calculations of Kimura et al~\cite{kimura} find a small ferromagnetic $\gamma^\prime$.

The fact that the value of $\gamma$ needed to get into this uudd region is now $>1$ might be discouraging. Also,
$a$ needs to be $\approx 1/2$, which also might not bode well for the present mechanism. However, the unoccupied
Mn orbital ($e_g$) in the manganites gives rise to a ferromagnetic contribution  to the Heisenberg exchange in
addition to the usual antiferromagnetic contribution.~\cite{goodenough} The resulting cancellation can be large
if the unoccupied orbital lies close in energy to the occupied orbitals, with the biquadratic exchange not
suffering such cancellation.~\cite{mila} And the Mn ion in the manganites apparently satisfies this requirement.
This close cancellation has been invoked for the nn exchange in a different mechanism for the origin of
uudd.~\cite{zhou} It has also been invoked to justify very large anisotropies compared to
$|J_1|$~\cite{mochizuki, mochizuki2}. But the latter, particularly the Dzyaloshinskii-Moriya interaction, is
expected to be $<<$ the antiferromagnetic term, being $\approx (g-2)/g$ times that term~\cite{moriya} (e.g., in
LaMnO$_3$, this is 1\%\cite{alejandro}, compared to the 10's of \% for the biquadratic terms). In this light, a
mechanism along the present lines (i.e. involving isotropic corrections to Heisenberg interactions) is clearly a
strong candidate for the origin of the uudd state in manganites. The Ferro (which leads to the A-type
ordering~\cite{mochizuki}) and Spiral regions also essentially account for the other ground state orderings
observed. The existence of spirals appears to rule out the non-frustrated model~\cite{zhou} as a general theory.

\textbf{Case 2. $\gamma=\gamma^\prime$, square symmetry.}\\
Under the nn interaction $J_1\rightarrow -J_1$, the net spin in the CF$_2$ and CF$_4$ remains non-zero, although
at a smaller value. Interestingly, this net spin occurs despite having only antiferromagnetic interactions in a
Bravais lattice. Uniform rotation by $\pm\pi/2$ of the horizontal arrows in the VL state in FIG. 4 changes it to
one of the $(\pi,0),(0,\pi)$ states of FIG. 2. At $a=0$, such a uniform rotation through an arbitrary angle
$\phi$ has energy independent of $\phi$ for any $\gamma$~\cite{henley2,kubo}, explaining why the boundary
between VL and $(\pi,0),(0,\pi)$ is the line $a=0$.

The question of what removes the degeneracy was considered: Randomness due to dilution was found to give
preference to $\phi=0$~\cite{henley3,henley2,henley} while quantum fluctuations stabilize $\phi=\pm\pi/2$, i.e.
the collinear states $(\pi,0),(0,\pi)$~\cite{henley2, kubo}. Furthermore, as we have seen, the same effect is
caused by the biquadratic terms, illustrating the use of the latter to mimic the fluctuations~\cite{henley,
nikuni}. In view of the appreciable size of the biquadratic terms, shown by experiment~\cite{harris,rodbell},
true biquadratic interactions might be at least as important as the fluctuations.

The purely electronic mechanism for the (2-body) biquadratic terms also gives, in the same order in the hopping
amplitude, 3-body, e.g. $\Sb_1\cdot\Sb_2\Sb_2\cdot\Sb_3$, and 4-body terms, like
$\Sb_1\cdot\Sb_2\Sb_3\cdot\Sb_4.$ To be complete one needs information about the coefficients of these various
terms, particularly their signs. The only unambiguous experiments, in that they can contain only 2-body terms,
are studies of magnetic dimers. Two examples: Mn impurities in MgO~\cite{harris}, where Mn-Mn pairs were
studied, and an example involving Ni$^{2+}$ dimers~\cite{bastardis2}. In the former case $a
>0$, in the latter $a<0$. Understanding of how either sign can occur can be seen in the perturbation
calculation of Bastardis et al~\cite{bastardis}. Unfortunately, such a conclusive result is not available for
the 3- and 4-body terms, as far as we're aware.  There is a calculation of the 3-body terms for a rather special
case~\cite{bastardis}, and the 4-body terms have been calculated only for $S=1/2$
spins~\cite{takahashi,chubukov}. The lattice-induced mechanism is similar in that it also gives 4-body
terms~\cite{wagner}, and sufficiently general explicit calculations of these terms are not available.
Fortunately, the experiments on MnO, NiO~\cite{rodbell}, where these extra terms will appear, show the same
physics as represented by the biquadratic terms with $a>0$, namely a preference for collinearity, thus a
stiffening of the collinear antiferromagnetic state. I.e., the extra terms do not necessarily spoil the reason
for the existence of the uudd or E-type state in our model. Thus we feel that the mechanism presented here for
the uudd state is probably correct.

%It would of course be of interest to see experimentally the exotic spin states, CF and conical vortex state.
%Recently the apparently first experimental study of the order by disorder effect has been reported~\cite{cheong}
%on a system where the disorder is random substitution of one magnetic ion for another, which should be
%essentially the same as the dilution discussed by Henley~\cite{henley2,henley3,henley} for the ordering.
%Unfortunately the conical vortex state that would be predicted cannot occur in this
%example because there apparently is strong Ising anisotropy.

In summary, we have shown that an essentially realistic model for the insulating manganites (the rhombic case)
captures the main ground-state magnetic features seen in these materials, spirals, A-type  and uudd or E-type
ordering. Isotropic corrections to frustrated Heisenberg interactions, in the simplified form of biquadratic
terms, characterize the model, a square symmetry version having also been studied. And, despite the model's
complexity, the LK cluster method~\cite{lyons} has been shown to enable simple and exact determination of the
classical ground states. Finally, the square symmetry case shows a novel spin ordering, the conical vortex
lattice, which might be accessible in real materials.

We thank C. Henley, C. Piermarocchi, A. Chubukov, J. B. Goodenough, M. Mochizuki for helpful discussions, and M.
Dykman and A. Kamenev for encouragement.

\vspace{.1in}
 \textbf{APPENDIX}

This contains some comments explaining further the basic cluster method, the derivations of the ground spin
states, and of statements about the degeneracy of various states.\\

\textbf{Additional explanatory remarks concerning the cluster method}\\
At the urging of referees, we add some hopefully clarifying remarks about the LK cluster method (despite these
having been made rather extensively in the original paper (1964)). We first note that there is an infinite
number of ways of writing $H$ of (1) in terms of cluster energies. (1) itself is one way, where each cluster is
either a nearest-neighbor pair of spins, a pair of (1,1)-diagonal nnn's, or a (1,-1)-diagonal nnn pair; and one
sums over all these ``clusters" precisely as written in (1). One will readily see that the ground state of the
individual clusters will not propagate when there are competing interactions, i.e. frustration. E.g., if $J_1<0$
and $A>0$, then the minimum for every nn-pair cluster will force the spins in each such pair to be parallel; but
given $J_2>0$, (1,1)-pair clusters will be minimized with antiparallel spins (assuming $A\ge0$). Thus the
essential idea of the method is to see if there is a more judicious choice. Experience has suggested that
highly-symmetric clusters have a much better chance of producing cluster ground states that propagate. Thus the
spins on a square plaquette were chosen in this case. The next step is to choose the cluster energy such that
the sum over \emph{every} such cluster (i.e. every plaquette) returns the original Hamiltonian, i.e. we need to
prove (2) for our choice (3). Substituting the nn pairs along $\hat{x}$ in (3) (from n=1 and n=3) into the right
side of (2) we have all the horizontal bonds
\begin{eqnarray}
H_x&=&-(1/2)\{\sum_\nb[\Sb_\nb\cdot\Sb_{\nb+\hat{x}}+a(\Sb_\nb\cdot\Sb_{\nb+\hat{x}})^2]+\nonumber\\ &
&\sum_\nb[\Sb_{\nb+\hat{y}}\cdot\Sb_{\nb+\hat{y}+\hat{x}}+a(\Sb_{\nb+\hat{y}}\cdot\Sb_{\nb+\hat{y}+\hat{x}})^2]\nonumber\}.
\end{eqnarray}
Changing the summation variable $\nb$ to $\nb-\hat{y}$ in the second sum leads precisely to the first sum,
together they cancel the 1/2, giving precisely the horizontal-bond terms in (1). The nnn Heisenberg bonds along
(1,1) come from the terms in $H_c$ in (2) of the form $\Sb_\nb\cdot\Sb_{\nb+\hat{x}+\hat{y}}$, and summing these
over all $\nb$ gives directly all the corresponding terms in (1), etc.\\
\textbf{Derivation of the macroscopic ground states via the
cluster method}\\
As seen from equations (2) and (3), the 4 spins in a cluster are labelled 1,2,3,4 going counterclockwise around
the square (the x and y directions are to the right and up, respectively). For coplanar states, $h_c$ depends
only on the angles $\theta_2,\theta_3,\theta_4$, of spins 2,3, and 4 relative to spin 1:
\begin{eqnarray}
h_c&\equiv& h(\theta_2,\theta_3,\theta_4)\nonumber\\
&=&-(1/2)(\cos\theta_2+\cos\theta_{23}+\cos\theta_{34}+\cos\theta_4)\nonumber\\
&&-(a/2)(\cos^2\theta_2+\cos^2\theta_{23}+\cos^2\theta_{34}+\cos^2\theta_4)\nonumber\\
&&+\gamma\cos\theta_3+\gamma^\prime\cos\theta_{24},\nonumber
\end{eqnarray}
where $\theta_{nm}=\theta_n-\theta_m.$ These states are denoted ($\theta_2,\theta_3,\theta_4$), and are
discussed first (in Cases 1 and 2 below). The procedure is to determine stationary states analytically,
solutions of $\frac{\partial h_c}{\partial\theta_n}=0$, see that they propagate, compare their energies, and
create a tentative ground state phase diagram. We then check that no lower cluster states were missed by various
numerical and other methods. For clarity, we first discuss the initial cluster states and their propagation into
crystal states, assuming our tentative phase diagram is correct. See the last section of the Supplement for
discussion of the checks made.

We will often refer to states related by symmetry (giving rise to ``trivial degeneracy" in Henley's terms (ref.
[26] main text)) as ``a state" and to states not related by symmetry as ``distinct states".

\textbf{Case 1. Infinitesimal $\gamma^\prime$ (rhombic symmetry)}\\
  For $\gamma^\prime=0$, in the region of FIG. 1 labelled Ferro, the minimum $h_c$ occurs for the state (0, 0, 0).
  In the region
  uudd,$(\pi,0)/(0,\pi)$
  the cluster ground states are $(\pi,\pi, \pi)$, $(\pi,\pi,0$) and ($0, \pi,\pi$). Taking $\Sb_1$ up, these can
  be written uddd, uddu, and uudd. The first, uddd, and its symmetry equivalents duuu, uudu, and ddud (since
  $\gamma^\prime\ne\gamma$, uuud and uduu are not equivalent to uddd), can be seen to propagate in the crystal state
  labelled uudd in FIG.2, establishing this state as a ground state (in the TL). The symmetry equivalent
  cluster states uddu and duud are seen to propagate in the $(\pi,0)$ state (on the left in FIG.2), the one on
  the right $(0,\pi$) comes
  from the uudd and dduu cluster states. The fact that all three cluster states are degenerate can be
  seen by inspection of FIG. 2 (the nn Heisenberg contribution is zero, the nnn contribution is the same for
  every plaquette). The degeneracy  between the uudd and $(0,\pi)$ is removed by $\gamma^\prime \ne0$,
  seen by inspection of FIG.2

  In the spiral region of FIG. 1, the lowest
cluster state for $\gamma>0$ is $(q_0, 2q_0, q_0)$, $\cos q_0 = (2\gamma-2a)^{-1}$.  From the uniform spin
rotation invariance of $h_c$, this is seen to propagate as a simple spiral, with wave vector $\mathbf{q} = (q_0,
q_0)$.

In the ``CF$_2$-Spiral(1,-1)" region, which occurs at $\gamma < 0$, $a<-1/2$, there are two degenerate cluster
ground states. One is $(q_1,0,q_1)$ ($\cos q_1=-1/(2a))$, pictured in FIG.Ba. It propagates uniquely into the
canted state CF$_2$ (FIG. 4). The other is $(q_1,0,-q_1)$, (Fig. Bc), which is seen to propagate as a spiral
with wave vector $\qb=(q_1,-q_1)$. But propagation can involve both these cluster states, leading to large
degeneracy, as discussed further below. The $\gamma$-independence of $q_1$ is an obvious consequence of spins 1
and 3 always being parallel for any $\qb$ in the (1,-1) direction.
This parallelism explains the $\gamma$-independence of the Spiral/CF$_2$-Ferro boundary.\\

\textbf{Case 2. $\gamma=\gamma^\prime$ (square symmetry)}\\
   In the regions of FIG.3 (XY and HEIS) labelled $(\pi,0)/(0,\pi)$, the cluster ground state is
$(\pi,\pi,0)$ = uddu (plus its symmetry equivalents), which, as we just saw, leads to $(\pi,0),(0,\pi)$ shown in
FIG. 2.

   In the VL (vortex lattice) regions,
minimum $h_c$ occurs for $(- \pi/2, \pi, \pi/2)$ and its symmetry equivalents. It is convenient to consider the
particular equivalent states obtained by reflection $\sigma_h$ of the spin positions in the horizontal line or
$\sigma_v$ in the vertical line (symmetry operations of $h_c$ in Case 2):
\begin{eqnarray}
\sigma_h\left( \begin{array}{cc} 4&3\\
1&2\end{array}\right)&\equiv&\left( \begin{array}{cc} 1&2\\
4&3\end{array}\right)\nonumber\\
\sigma_v\left( \begin{array}{cc} 4&3\\
1&2\end{array}\right)&\equiv&\left( \begin{array}{cc} 3&4\\
2&1\end{array}\right).\nonumber
\end{eqnarray}
Also define $T_x,T_y$ as translations through a lattice constant in the $x,y$ directions respectively. Applying
$T_x\sigma_v$ successively to the plaquette in the lower left of FIG. 4 VL, then applying $T_y\sigma_h$
successively to that result, and so on, one sees that the whole figure is reproduced. (This is a series of
checker moves, moving a column (or row) over the other column (or row), but not removing the ``jumped" spins.)
Hence every plaquette has minimum $h_c$ so that VL is a crystal ground state in this region. Essential to this
propagation is the commutation, $T_y\sigma_h T_x\sigma_v=T_x\sigma_v T_y\sigma_h$, giving the 4th plaquette (the
central one in the figure) the same for each possible path to it.

These considerations lead directly to the following: \emph{Any} set of 4 cluster spins propagates in this way
for square symmetry. Thus \emph{the cluster method rigorously reduces the $N$-spin problem to a 4-spin problem
for any square-symmetric interactions which can be described in terms of the square plaquette clusters}.

   In the region labelled ``Spiral, CF$_4$", the cluster ground state is the same, $(\theta,2\theta,\theta)$
   with $\theta = q_0$, as in
   the spiral region of FIG. 1. This can propagate as a spiral with wave vector $\qb=(q_0,q_0)$, or its symmetry-caused
   degenerate
   counterpart, the spiral with wave vector $(q_0, -q_0)$ (from cluster state $(\theta,0,-\theta))$, as well as the spirals
   with $\qb\rightarrow -\qb$.
   However, surprisingly, there is more than one way that this cluster state can propagate, one of which is
   the 4-sublattice canted ferromagnet CF$_4$ shown in FIG. 4, which comes from propagating by repeated application of
   $T_x\sigma_y$ and $T_y\sigma_x$ to the basic cluster $(\theta,2\theta,\theta)$. In fact there is a
   large number of degenerate states, discussed below.

   One can view these different propagations generally as
   applying a lattice translation $T_\nb$ times a symmetry operation
   of the cluster. For the spiral the cluster symmetry operation is a uniform spin rotation $R_\theta$;
   in the CF$_4$ case the
    cluster symmetry operation is either $\sigma_v$ or $\sigma_h$. Since in this case these operations are seen to
     yield no contradiction, (again,essentially because $[\sigma_v,\sigma_h]=0$),
     the CF$_4$ state is established as a ground state. The other degenerate states come from applications of $T_\nb$
     times one or the other of $R,\sigma_v,\sigma_h$.

     Note that large degeneracy of crystal states has originated from degeneracy of two distinct cluster states (in the
     rhombic symmetry case), whereas for square symmetry, it came from different propagations of a
     single (symmetry-induced) cluster state.\\
\begin{figure}[h]
\includegraphics[height=3in]{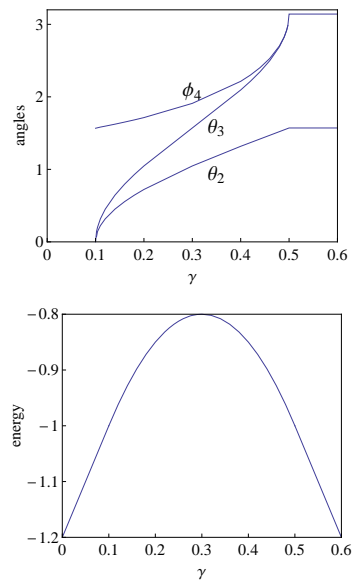}
 \caption{Variation of angles and energy with $\gamma$ at $a=-0.4$. The non-coplanar boundaries are at
 $\gamma=0.1$ and 0.5, with Ferro for $\gamma<0.1$ and VL for $\gamma>0.5$. Energy vs. $\gamma$ is linear in the
 latter regions.}
 \label{suppfigA}
\end{figure}
     \textbf{Non-coplanar states}
     To examine the possibility that the Heisenberg ground state is not coplanar, we calculated $h_c$ over a mesh
     with 5 angles varying independently (polar angles $\theta_n$ for n=2,3,4, azimuthal angles $\phi_n$ for
     n=3,4; $\Sb_1\equiv\hat{z})$. This is completely general due to the spin-rotational symmetry.
     Doing this at sample points in each of the regions of FIG.'s 1 and 3, we found
     instability with respect to deviation from coplanarity only in the region of FIG.3HEIS labelled ``Conical VL".
     We then did a closer
     examination as follows. We analytically found
     instability on the boundary between Spiral,CF$_4$ and VL in FIG. 3XY.
     We then used the $\theta-$values of the spiral (or CF$_4$) at a point on this boundary as an estimate in the
     FindMinimum program of Mathematica to determine a nearby minimum of $h_c$. We then repeated the
     calculation of FindMinimum at neighboring points thus generating the ground state over the phase diagram, yielding
     FIG. 3HEIS. The results revealed the general property $\theta_2=\theta_4, \phi_3=\phi_4/2$.
     We found that the states approached the VL state on the vertical boundary $\gamma=1/2$, and
     the ferromagnetic state on the same line $a=\gamma-1/2$ as the Ferro-Spiral,CF$_4$  boundary in FIG.3. A sample
     behavior of the angles and energy as $\gamma$ varies with fixed $a$ is shown in FIG. A.

     Looking at the example non-coplanar state in FIG. 5, we saw no symmetry at all. This seemed strange in
     view of the very simple boundary structure found (FIG.3HEIS). After much puzzling over this aesthetically unsatisfying
     situation, we realized that there is
     a very simple picture of the non-coplanar state! From the numerically-determined cluster state,
     we found the scalar products of all 4
     nn spins to be equal. This implies that the spins in a single plaquette lie on the surface of
     a cone, 1/2-angle $\Omega$, with equally spaced azimuthal angles $\phi$, i.e. the nn $\phi$ spacing is
     $\pi/2$. This is described by
     \begin{eqnarray}
     \Sb_n&=&\sin\Omega(\hat{x}\cos n\pi/2+\hat{y}\sin n\pi/2)+\cos\Omega\ \hat{z},\nonumber\\
     & &n=1,\cdots,4.\nonumber
     \end{eqnarray}
    The energy $h_c$ is now easily written down:
     \begin{equation}
     h_{CVL}(\Omega)=-2\cos^2\Omega-2a \cos^4\Omega+2\gamma(2\cos^2\Omega-1).\nonumber
     \end{equation}
     The projection of the spins on the x-y plane propagates to exactly the vortex lattice with reduced spin lengths; thus
     the name ``Conical VL" (CVL).   For $a<0$, $h_{CVL}$ is minimum at
     \begin{equation}
     \cos^2\Omega=(2\gamma-1)/(2a)\equiv\cos^2\Omega_0\nonumber
     \end{equation}
     for $0\le(2\gamma-1)/(2a)\le 1$, with corresponding energy
     \begin{equation}
     h_{CVL}=(1-2\gamma)^2/(2a)-2\gamma.\nonumber
     \end{equation}
     The other cluster energies relevant to FIG. 3HEIS are
     \begin{eqnarray}
     h_{Ferro}&=&-2-2a+2\gamma\nonumber\\
     h_{(\pi,\ 0)}&=&-2\gamma-2a\nonumber\\
     h_{VL}&=&-2\gamma\nonumber\\
     h_{CF_2}&=&1/(2a)+2\gamma.\nonumber
     \end{eqnarray}
     It is readily verified that these equations yield the boundaries in FIG.3HEIS, those bounding the CVL
     region having previously been determined numerically.
    It is also seen that the cluster state $\Sb_n,n=1\cdots 4, \rightarrow$ the Ferro state as $\Omega\rightarrow0$ and
    the VL state as
    $\Omega\rightarrow\pi/2$. This implies continuous transitions at the respective
     boundaries (see also FIG.A). At the CF$_2$-CVL boundary, $\gamma=0,a<-1/2$, one checks that the energies are the same,
    but the spin states differ, implying a 1st order phase transition.

\textbf{On the degeneracy in various regions.} \\
When $\gamma=0$, for either the rhombic or square case, there is a transition from ferromagnetism to a highly
degenerate ground state as $a$ decreases past -1/2. This occurs because at $a < -1/2$, the combination of nn
ferromagnetic Heisenberg and perpendicular-orientation-favoring biquadratic interactions requires an angle
between nn spins given by $\theta(=q_0)=\cos^{-1}\frac{-1}{2a}$. Thus for some direction of a given spin, its
nn's each are only restricted to lie on a cone of 1/2-angle $\theta$ measured from that spin. For simplicity we
consider XY spins, so the restriction is just to two relative directions $\pm\theta$.

For d=1 (TB), the degeneracy is asymptotically $2^N$:  given one spin, and moving in one direction, say to the
right, along the chain, its nn to the right has two possible directions, and for each of these, \emph{its} nn to
the right has 2 possible directions, etc. But for d=2, there are restrictions on the degeneracy of a pair of nn
spins depending on what the other nn's are, because of the loops that occur.
\begin{figure}[h]
\includegraphics[height=1.5in]{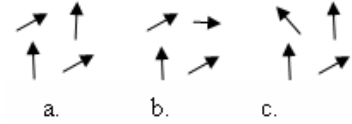}
 \caption{Degenerate plaquette states ($\gamma=0$).}
 \label{fig:suppfig1}
\end{figure}

Consider a square plaquette with its 4 spins making angles $\pm \theta$ with its nn's, and for ease of
visualization take $\theta=45\deg$. First fix two of them, say the bottom two; then there are 3 degenerate
states, shown in FIG.B. If one fixes 3 spins, then the situation is more complicated. If one fixes the 3 in the
lower left hand corner of Fig. Ba., then there are two possibilities for the 4th spin (as in a. and  b.). But if
the 3 spins are as in c., the 4th spin has only one possibility. It is this constraint that complicates the
counting. While it seems that there is probably ``macroscopic degeneracy", $\mathcal{S}$=O$(N)$, we have not
been able to show it, because of this constraint. We can however show that the degeneracy is at least that where
$\mathcal{S}$=O$(N^{1/2})$. If one considers one row, length $\sqrt{N}$, of the (2d) crystal, one can see that
any set of spins such that each spin makes angle $\pm \theta$ with its nn's, (as in the 1d case), is possible in
the ground state of the 2d crystal. For each of these it will always be possible to build a crystal ground state
(in the TL) by propagating the clusters into the 2nd dimension. Thus the number of ground states is at least
$2^{\surd{N}}$.

For $\gamma>0$ in the square lattice XY case, FIG.Ba is higher energy than b or c. Nevertheless, there is still
a large degeneracy, $\mathcal{S}$ at least O$(N^{1/2})$, seen by the same argument just given. In the region
uudd, $(\pi,0)/(0\pi)$, FIG. 1, an essentially similar argument gives the degeneracy at least of
O($2^{\surd{N}})$.

Also in the square symmetry case, when $\gamma<0$, only the state Fig.Ba is lowest; in this case propagation can
occur only through the reflections and leads uniquely to CF$_2$. For the case $\gamma^\prime=0,\gamma>0$, only
one cluster state, FIG.Bb, is lowest, so the bound is unity, and the only state is the spiral (the rhombic
symmetry removes the reflections $\sigma_v,\sigma_h$ as symmetry operations); the latter case is discussed in
more detail in the next section.

\textbf{Uniqueness of the cluster propagation in some regions.} \\
The existence of a huge number of different ways to propagate the given cluster ground states in some regions
forces investigation of a similar possibility in other regions.  We find unique propagation in the regions
Ferro, Spiral(1,1) (FIG. 1), and $(\pi,0)/(0\pi)$, VL, and Conical VL (FIG. 3). We give a proof in the case of
Spiral (1,1), illustrating the procedure used for the other cases.

To propagate a cluster state one must consider translations $T_x$ and $T_y$. But, as we've seen, there are
symmetries of the cluster states that can also be involved. One can see that if $\hat{O}$ is such a symmetry
operation, a necessary condition for propagation is
\begin{equation}
\hat{O} \Sb_1=\Sb_2\ \mbox{and}\ \hat{O}\Sb_4=\Sb_3.\nonumber
\end{equation}
This comes from the anticipated application of $T_x$. A similar condition occurs for $T_y$.

We have $(\theta,2\theta,\theta)\equiv\psi$ as the cluster state associated with the (1,1) spiral, wave vector
$\Qb=(q_0,q_0)$, $q_0>0$, and we are considering the case of rhombic symmetry. We confine the proof to XY spins.
The symmetry operation that yields the propagation into this spiral is $\hat{O}=R_\theta$, rotation of the four
spins by $\theta$, as already discussed. The question here is, ``Are there any other $\hat{O}$'s that will allow
a different propagation?". Fortunately, there is only a small number of possibilities, namely the spatial
operations of the rhombus, and those times some spin rotation or reflection applied to all four spins. The
rhombus operations are $\sigma_{1,1},\sigma_{1,-1}, \rho_\pi$, respectively, reflection in the two diagonals,
and rotation through $\pi$. Clearly $\sigma_{1,1}\psi=\psi$: no new information. Assume $\theta=\pi/4$ for
simplicity.

Writing $\psi=\left(\begin{array}{cc}\nearrow&\rightarrow\\\uparrow&\nearrow\end{array}\right)$, we have
$\sigma_{1,-1}\psi=\left(\begin{array}{cc}\nearrow&\uparrow\\\rightarrow&\nearrow\end{array}\right)$. There are
two possibilities to operate now with spin operation $\hat{O}_s$ to satisfy $\hat{O}\Sb_1=\Sb_2$,
$\hat{O}=R_\theta\ \mbox{or}\ \sigma_s$, respectively rotation through $\theta$ or reflection through the line
$y=(\tan\pi/8) x$.
$R_\theta\sigma_{1,-1}\psi\equiv{P}\psi=\left(\begin{array}{cc}\uparrow&\nwarrow\\\nearrow&\uparrow\end{array}\right)$,
showing that $P$ takes $\Sb_1$ to $\Sb_2$ (by design), but takes $\Sb_4=\nearrow$ into
$\uparrow\ne\Sb_3(=\rightarrow).$ So this path does not lead to propagation.

The other possibility, replacing $R_\theta$ by $\sigma_s$. We have
$\sigma_s\sigma_{1,-1}\psi=\left(\begin{array}{cc}\rightarrow&\searrow\\\nearrow&\rightarrow\end{array}\right)$,
which is just $R_\theta$, so nothing new.

The only remaining possibility (excluding $\rho_\pi$) is $\sigma_s^\prime$, reflection in the line
$y=(\tan3\pi/8) x$, applied directly to $\psi:
\sigma_s^\prime\psi=\left(\begin{array}{cc}\uparrow&\nwarrow\\\nearrow&\uparrow\end{array}\right)$. But this has
violated $\hat{O}\Sb_4=\Sb_3$. Interestingly, the last spin state would propagate as a spiral with wave vector
-$\Qb$. Finally we note that $\rho_\pi\psi=\sigma_{1,-1}\psi$, already considered. We can conclude that the
propagation in the spiral region of FIG. 1 is unique.

The reason we cannot conclude that a state is non-degenerate even if there is a unique propagation of the
cluster ground state is that we know only that the state so-obtained is \underline{a} ground state. This is
similar to the case of the Heisenberg Hamiltonian on a Bravais lattice: we know that the ground state energy is
necessarily obtained by the minimum-energy spiral or spirals (ref. 21). And while the spirals are usually the
only ground states, there are quite special cases where there are additional degeneracies. See e.g. Z. Nussinov,
cond-mat/0105253v12.

\textbf{Checks on the tentative ground states} \\
The most straightforward check is to consider a region where we suspect $h_s$ is the minimum and simply
calculate $h_c-h_s$ over a mesh that covers the full range of the (3 or 5) angle variables in $h_c$. Usually we
took the mesh step $\delta$ as $\pi/10$, reasonable in view of the fact that the most rapidly changing function
is $\cos 2\alpha$ where $\alpha$ is one of the angles (giving a ``length scale" of ~$\pi/2$). Some places we
used $\pi/20$ instead. This procedure checked all the regions. A slight problem occurred very near first-order
boundaries--quite understandable: even if the function is very well represented by the values on the mesh, if
two local minima are very close in energy, depending on how the mesh points fall, the true minimum might not be
found. This problem was completely overcome by using Mathematica's FindMinimum program, which searches for a
local minimum given a starting point P. We ran this with P on a mesh running over the full many-angle space.
Then for any point on the P-mesh that falls within the basin of a particular local minimum immediately goes to
that minimum value, with arbitrary precision. The required interval for this mesh $\delta_P$ is not as tight as
$\delta$. As an example, using this more powerful method, the vertical (1st-order) boundary at $\gamma=0$ was
preserved to within one part in $10^4$ or better, using $\delta_P=\pi/4$.

Another check was with Mathematica's program Reduce, which analytically is supposed to return all the solutions
to the stationarity equations. This worked for some regions in the sense that it ran in short time (~few
minutes), but in other regions it ran for at least hours, and we didn't wait. Where it did work, it confirmed
our initial results, giving a rigorous proof for those regions.

We also note that the ground state energy is rigorously known on the lines $a=0$ and $\gamma=0$, the former by
the Luttinger-Tisza method, the latter by the cluster method, where the clusters are just the 2-body terms in
the original form of the Hamiltonian. Also the limit $a\rightarrow\infty$ is clearly correct, as well as the
limit $\gamma\rightarrow-\infty, a\rightarrow-\infty$.

These considerations have convinced us that our analytically-described phase diagrams are exact, although we
can't claim a rigorous proof due to the use of these numerical methods as checks.

%\begin{thebibliography}{99}
%\bibitem{lyons}D. H. Lyons and T.A. Kaplan, J. Phys. Chem. Solids \textbf{25}, 645 (1964); Erratum, \emph{ibid.}
%pg. 1501.

%\end{thebibliography}

\begin{thebibliography}{99}
\bibitem{ref} We neglect anisotropies which remove degeneracies and cause small
distortions of spin
states.
\end{thebibliography}
\end{document}